\documentclass[conference,twocolumn,10pt,letterpaper]{IEEEtran}

\usepackage[cmex10]{amsmath} % cmex10 proposed by IEEEtran_HOWTO
\usepackage{amssymb,amsthm}

\usepackage{tikz}
\usetikzlibrary{calc}
\usepackage{pgfplots}
\usetikzlibrary{positioning}

\usepackage{calc}

\newcommand{\fm}[1]{\scriptsize\mbox{\ensuremath{#1}}}
\newcommand{\ft}[1]{\scriptsize #1}

\newtheorem{theorem}{Theorem}

\newtheorem{corollary}{Corollary}

\newcommand{\ie}{{\it i.e.},\ }

\newcommand{\FF}{\mathbb{F}}

\newcommand{\RR}{\mathbb{R}}

\newcommand{\LL}{\mathcal{L}}

\newcommand{\Rnet}{R_{\mathrm{net}}}
\newcommand{\barRnet}{\overline{R}_{\mathrm{net}}}
\newcommand{\Rresolve}{R_{\mathrm{res}}}
\newcommand{\barRresolve}{\overline{R}_{\mathrm{res}}}

\newcommand{\Rplnc}{R_{\mathrm{plnc}}}

\begin{document}

\title{Sign-Compute-Resolve for Random Access}

\author{
\IEEEauthorblockN{
Jasper Goseling\IEEEauthorrefmark{1}\IEEEauthorrefmark{3},
\v Cedomir Stefanovi\' c \IEEEauthorrefmark{2} and
Petar Popovski\IEEEauthorrefmark{2} 
}\\
\IEEEauthorblockA{
 \IEEEauthorrefmark{1}
 Stochastic Operations Research, University of Twente, The Netherlands\\
}
\IEEEauthorblockA{
 \IEEEauthorrefmark{2}
 Department of Electronic Systems,
 Aalborg University, Denmark
}
\IEEEauthorblockA{
 \IEEEauthorrefmark{3}
 Delft University of Technology, The Netherlands\\
 j.goseling@utwente.nl, cs@es.aau.dk, petarp@es.aau.dk
}
}

\maketitle

%%%%%%%%%%%%%%%%%%%%%%%%%%%%%%%%%%%%%%%%%%%%%%%%%%%%%%
%
%
%
%%%%%%%%%%%%%%%%%%%%%%%%%%%%%%%%%%%%%%%%%%%%%%%%%%%%%%
\begin{abstract}
We present an approach to random access that is based on three elements: physical-layer network coding, signature codes and tree splitting. Upon occurrence of a collision, physical-layer network coding enables the receiver to decode the sum of the information that was transmitted by the individual users. For each user this information consists of the data that the user wants to communicate as well as the user's signature. As long as no more than $K$ users collide, their identities can be recovered from the sum of their signatures. A splitting protocol is used to deal with the case that more than $K$ users collide. We measure the performance of the proposed method in terms of user resolution rate as well as overall throughput of the system. The results show that our approach significantly increases the performance of the system even compared to coded random access, where collisions are not wasted, but are reused in successive interference cancellation. 
\end{abstract}

%%%%%%%%%%%%%%%%%%%%%%%%%%%%%%%%%%%%%%%%%%%%%%%%%%%%%%
%
% Introduction
%
%%%%%%%%%%%%%%%%%%%%%%%%%%%%%%%%%%%%%%%%%%%%%%%%%%%%%%
\section{Introduction} \label{sec:intro}
% !TEX root = signatures_main.tex

%%%%%%%%%%%%%%%%%%%%%%%%%%%%%%%%%%%%%%%%%%%%%%%%%%%%%%
%
% Introduction
%
%%%%%%%%%%%%%%%%%%%%%%%%%%%%%%%%%%%%%%%%%%%%%%%%%%%%%%

Uncertainty is the essential element of communication systems. In an information-theoretic setting, uncertainty is associated with noise, while in the context of communication protocols, uncertainty is associated with traffic (packet) arrivals at the users. A canonical example of the latter is seen in random access protocols, used for handling transmissions of users to a common receiver, e.g., a base station, over a shared wireless medium. 
Random access is necessary when the total number of users associated with the base station is very large, but at a given short time interval, the number of active users that have packets to transmit is small and a priori not known.
Such is the case for, for instance, wide-area networks of sensors, where each sensor has a sporadic traffic pattern.
The goal of random access protocols is to enable each of the active users to eventually send her packet successfully. 

Traditionally, random access protocols have been designed under the \emph{collision model}: when two or more users transmit at the same time, a collision occurs and all involved transmissions are lost.
In other words, collisions are considered destructive and the information that is contained in them as irrecoverable.
Therefore, the objective of classical random access protocols, such as ALOHA \cite{R1975} or splitting tree \cite{capetanakis}, is to ensure that each user gets the opportunity to send its packet without collision.
Recently, a generalization of the collision model, obtained by including a more elaborate physical-layer model, brings in the possibility for Successive Interference Cancellation (SIC) and gives rise to a new class of protocols, termed coded random access \cite{PSLP2014}.
The main feature of this model is that a collision is treated as a sum of packets and, instead of being discarded, it can be buffered and reused in a SIC-based decoding.
We illustrate this through a simple example, in which the received signals in the  first two slots are:
\begin{align}
\label{eq:SimpleSIC}
Y_1 & = X_1 + X_2 + Z_1, \nonumber \\
Y_2 & = X_2 + Z_2,
\end{align}
where $X_1$ and $X_2$ are the two useful signals (packets) and the $Z_1$, $Z_2$ are noise signals. 
The received signal $Y_1$ is buffered, and, if $X_2$ is successfully decoded from the singleton slot $Y_2$, it can be subtracted (i.e., cancelled) from $Y_1$, effectively reducing the first slot to a singleton.
The receiver proceeds by attempting to decode $X_1$ from the noisy signal $ Y_1 - X_2 = X_1 + Z $.

An important constraint of coded random access is that the useful signals must carry embedded pointers that inform the receiver where their replicas occurred.
In the above example, the replica of $X_2$ in $Y_2$ has a pointer that indicates that another replica occurred in $Y_1$, so that the receiver, after decoding $Y_2$, also learns that a replica of $X_2$ can be cancelled from $X_1$.
Another important aspect of coded random access protocols it that the receiver buffers analog signals that contain noise. Hence, the uncertainty brought by the noise persists while the protocol resolves the uncertainty about the set of active users. 

This is fundamentally changed by applying the ideas of Physical Layer Network Coding (PLNC) to the problem of random access.
The key idea in PLNC is to decode a function of multiple received signals, rather than decoding the individual signals. Such operation is termed denoise-and-forward (DNF) \cite{popovski2006anti} or compute-and-forward (CF) \cite{nazer11compforw}.
Let us reuse the example (\ref{eq:SimpleSIC}) and assume that $W_1$ and $W_2$ represent the data bits that are mapped to the baseband signals $X_1$ and $X_2$, respectively.
Upon receiving $Y_1$ from (\ref{eq:SimpleSIC}), the base station stores the bitwise XOR $W_1 \oplus W_2$. If $X_2$ $(W_2)$ is decoded from $Y_2$, then $W_1$ is recovered by XOR-ing $W_2$ with the stored signal $W_1 \oplus W_2$.
We can, therefore, say that the use of DNF (CF) removes the uncertainty of the noise from the protocol and deals only with the uncertainty of the contending set of users. 

Another limitation of coded random access is that the receiver must wait until it successfully decodes a packet from a singleton slot in order to start and maintain its operation. For example, let the receiver get the following signals in the first three slots:
\begin{eqnarray}\label{eq:3SlotsReceptionSIC}
Y_1 & = & X_1 + X_2 + Z_1, \nonumber \\
Y_2 & = & X_1 + X_3 + Z_2, \nonumber \\
Y_3 & = & X_2 + X_3 + Z_3.
\end{eqnarray}
These are three (noisy) equations with three unknowns $X_1, X_2, X_3$ and, in principle the receiver should be able to recover all three signals by using the same techniques used to decode multiple streams in a MIMO transmission~\cite{tse2005fundamentals}. However, recall that in a setup with random access, the receiver has no knowledge about the set of transmitting users, e.g., in this case it does not know that $X_1$ is sent in slots $1$ and $2$, $X_2$ in slots $1$ and $3$, etc. It therefore needs to wait to receive, e.g., $Y_4 = X_1 + Z_4$, decode $X_1$ and learn from the embedded pointers in which other slots had $X_1$ been sent, such that it can be canceled.
This fundamental limitation of SIC-based random access sets the motivation to introduce PLNC-based random access with signatures \cite{goseling14massap}.
For example (\ref{eq:3SlotsReceptionSIC}) and the simplest case of $\mathbb{F}_2$ functions, in PLNC-based random access the receiver stores the following three (noiseless) digital binary signals: 
\begin{eqnarray}\label{eq:3SlotsReceptionPLNC}
V_1 &=& W_1 \oplus W_2, \nonumber \\
V_2 &=& W_1 \oplus W_3, \nonumber \\
V_3 &=& W_2 \oplus W_3.
\end{eqnarray}
In this way the receiver observes a noiseless XOR multiple access channel.
The main idea of random access using PLNC and signatures is that the $\ell-$th user applies the following communication strategy: she prepends a \emph{signature} $W^s_\ell$, consisting of predefined number of bits, to the pure data $W_\ell^d$ in order to obtain $W_\ell$. The signature is based on a code that has the following property: if at most $K$ users transmit in a given slot, then from the \emph{integer sum} of the signatures $W^s_1 + W^s_2 + \cdots + W^s_L$,\footnote{This integer sum is obtained from  the PLNC output~$\bigoplus_{\ell = 1}^{L} W^s_{\ell}$, by making use of a result by Nazer~\cite{nazer2012successive}.} $L \leq K$, the receiver knows exactly which transmitters have contributed to the XOR-ed data stored in the present slot. In other words, the sum:
\begin{align} \label{eq:SignatureSum}
\sum_{\ell = 1}^{L} W^s_{\ell}
\end{align}
is uniquely decodable if $L \leq K$.
Referring to the example (\ref{eq:3SlotsReceptionPLNC}), the receiver will be able to decode the individual data already after the third slot.
The receiver also detects if the number of users sending in a slot is larger than $K$, and the stored XOR combination cannot be used for decoding based on signatures. It may, however, still be used further in the decoding process, as detailed in the sequel.

In this paper we leverage the idea of PLNC-based random access and design a Contention Resolution Algorithm (CRA) that uses signatures.
In contrast to collision avoidance protocols \cite{R1975}, contention resolution protocols \cite{capetanakis,Massey, popovski2007class} are efficient in terms of resolving collisions when they occur.
The conventional contention resolution algorithms drive the set of contending users towards the state in which each user gets the opportunity to transmit without interference from the others.
On the other hand, the use of signatures and PLNC generalizes the concept of collision by allowing the receiver to have metadata (\ie knowledge of the set of colliding users) about the observed collision.
This feature fundamentally changes the objective of a CRA: the set of contending users should be driven in a state where the receiver gets a sufficient number of \emph{equations} in the finite field in order to be able to decode the users' data.
We provide details on the basic tradeoffs and mechanisms that needs to be considered for a CRA based on PLNC and  signatures.
The results show that the use of signatures is significantly reducing the average time required to extract useful information from the collisions and therefore improve the overall throughput of the system.

The idea of using SIC in framed ALOHA setting was first proposed in \cite{CGH2007}.
The analogies of SIC-based ALOHA with erasure-coding theory were identified in \cite{liva2011graph}, establishing the paradigm of the coded random access that was further developed in \cite{PLC2011,stefanovic12frameless, LPLC2012}.
It was shown that coded random access achieves throughputs that asymptotically tend to 1.
The highest non-asymptotic throughputs were, so far, reported in \cite{SP2013}; e.g., when number of contending users is 1000, the expected throughput is 0.88.

The use of SIC in the contention resolution framework was first investigated in \cite{SICTA}.
Here it was shown that enhancing the original tree-splitting scheme \cite{capetanakis} with SIC doubles the asymptotically achievable throughput to 0.693.
Another approach was suggested in \cite{SSP2013}, where SIC was employed over a set of partially split trees, and optimization was performed over the splitting strategy that favors fast SIC evolution. The reported throughputs for the presented design example in \cite{SSP2013} are close to 0.8.

Finally, the use of PLNC for random access was studied in~\cite{parandehgheibi2010collision,parandehgheibi2010acknowledgement,cocco2011vtc,goseling2013random,goseling2013physical,cocco2014network} in which it was assumed that the receiver knows which users are active in each slot. The use of physical-layer network coding and signature codes was considered in~\cite{censor2012bounded} for broadcast in networks. The combination of physical-layer network coding and signature codes for random access was introduced in~\cite{goseling14massap}. In~\cite{censor2012bounded} as well as~\cite{goseling14massap} it was assumed that the number of contending users is bounded. In the current work we leverage this assumption and design a CRA that can deal with any number of contending users by incorporating tree splitting.

The paper is organized as follows. In Section~\ref{sec:model} we introduce our model. In Section~\ref{sec:prelim} we present some results on PLNC, signature codes and tree splitting that will be used in the remainder. The proposed strategy is presented in detail in Section~\ref{sec:idea}. The performance of the strategy is analyzed in Section~\ref{sec:analysis}.
The discussion and concluding remarks are given in Section~\ref{sec:discussion}.

%%%%%%%%%%%%%%%%%%%%%%%%%%%%%%%%%%%%%%%%%%%%%%%%%%%%%%
%
% Model
%
%%%%%%%%%%%%%%%%%%%%%%%%%%%%%%%%%%%%%%%%%%%%%%%%%%%%%%
\section{Model and Problem Statement} \label{sec:model}
% !TEX root =  signatures_main.tex

%%%%%%%%%%%%%%%%%%%%%%%%%%%%%%%%%%%%%%%%%%%%%%%%%%%%%%
%
% Model
%
%%%%%%%%%%%%%%%%%%%%%%%%%%%%%%%%%%%%%%%%%%%%%%%%%%%%%%

%%%%%%%%%%%%%%%%%%%%%%%%%%%%%%%%%%%%%%%%%%%%%%%%%%%%%%
We consider a system that has a total of $M$ devices (users).
Each of the users is assigned a unique identity from the set $\{1,\dots,M\}$.
For notational convenience in the remainder we assume that $M$ is prime.
Each user sporadically gets a data packet that needs to be sent to a receiver that is common for all users.
The users that have data to transmit wait for a beacon sent by the common receiver, which marks the start of the contention process.
In our model, the probability that a user has a packet to transmit when the beacon is sent is $p$, where $p$ is rather small, \ie $pM < <M$. 

Let $\LL$ denote the set of contending users, \ie the set of users that have a packet to transmit, and let $L=|\LL|$. The receiver does not know  $\LL$, otherwise the contention problem would have been trivial - the receiver would simply schedule the users from $\LL$. The scheduling can be based by asking the users to transmit with rates that correspond to a certain point within the achievable region of an $L-$dimensional multiple access channel. However, the receiver does not know $\LL$ and cannot make such a scheduling.
In some cases our interest will be in the performance conditioned on a number of active users $L = |\LL|$. Note that in that case only $L=|\LL|$ is of importance.
Since the packet arrivals across the set of users are independent, $L$ has a binomial distribution: the probability that $L$ users are active is denoted by $q(L) = P(|\mathcal{L}|=L) = \binom{M}{L}p^L(1-p)^{M-L}$. For notational convenience, let $q_0=q(0)=(1-p)^M$. We will be interested in the probability of having $L$ active users conditioned on the fact there is at least one. We denote this probability by $\hat q(L)$ and it readily follows that $\hat q(L) = P\left( |\mathcal{L}|=L \big{|} |\mathcal{L}|>0 \right) = q(L)/(1-q_0)$.  We will express some of our results in terms of $I_x(a,b)$, the regularized incomplete beta function, which is defined as $I(x;\ a,b)=B(x;\ a,b)/B(1;\ a,b)$ with $B(x;\ a,b) = \int_0^x t^{a-1}(1-t)^{b-1}dt$. The reason is that $\sum_{L=0}^K q(L) =  I_{1-p}(M-K,K+1)$. 

The data packet of each contending user consists of $D$ bits. The channel coefficient between the $m-$th user and the receiver is $h_m$. Due to reciprocity, the contending device is capable to  estimate the channel and \emph{precode} its transmission by transmitting the signal $\frac{X_m}{h_i}$. The time starts at $\tau=1$ and the $\tau-$th transmitted symbol by the $m-$th user is denoted by $X_m(\tau)$. Hence, at the $\tau-$th channel use, the receiver observes 
\begin{equation}
Y(\tau) = \sum_{m \in \LL} X_m(\tau) + Z(\tau),
\end{equation}
where $\LL$ is the set of active, contending users.
$Z(\tau)$ is the Gaussian noise with unit variance.
Each user transmits at the same rate (in bits/channel use) and one packet transmission has a duration of a \emph{slot} that consists of $N$ channel uses.
At the end of each slot, the common receiver provides feedback to the users. This feedback is instantaneous, error-free and received by all users.
We do not impose any constraints on the amount of feedback that can be provided; we will explicitly specify how feedback is used later in the paper.

Further, we will assume that the signal of each user needs to satisfy an average power constraint in each round, \ie
\begin{equation}
\frac{1}{N}\sum_{\tau=1}^{N} \left|X_m(\tau)\right|^2 \leq P,
\end{equation}
for all $m\in\{1,\dots,M\}$.
We will assume that $P>1$, such that the Signal-to-Noise Ratio (SNR) is also larger than one, which is required to have a nontrivial computation rate over the multiple access channel, as seen in the next section.
The reader may object that the actual transmitted power by the user can be much higher than $P$, since each user inverts the channel.
This can be addressed by assuming that a user that observes a channel with $|h_m|$ lower than a threshold, does not join the set of contending users; in that case the probability $p$ also accounts for the fact that the user channel is sufficiently strong, in addition to the assumption that the user has a packet to send. 

For simplicity, as it is common in PLNC schemes, we assume that the channel input/outputs are real, \ie $X_m(\tau) , Z(\tau), Y(\tau) \in \RR$. The results are readily transferable to the case of complex symbols, by doubling the number of bits per channel use.

The goal of this paper is to devise a protocol that allows the receiver to retrieve both the identities and the data packets of all active users.
The constituent elements of the protocol are the use of contention resolution mechanism across slots, dealing with randomness of the user activity pattern, and use of forward error correcting code within slots, dealing with noise.
With respect to the latter, we ignore finite block length effects and assume that forward error correcting codes operate with zero error at any rate up to and including capacity.
As a consequence, the task for the receiver is to recover all packets with zero error probability.

We are interested in the following performance parameters.
By $S(L)$ we denote the expected number of slots that the protocol uses to resolve $L$ contending users, where the expectation in $S(L)$ is w.r.t.\ the randomness in the contention resolution mechanism.
By $\Rresolve(L) = L/S(L)$ we denote the expected number of users that is resolved per slot.
We are also interested in $\barRresolve$, obtained by averaging $\Rresolve(L)$ over $L$, \ie 
\begin{equation}
 \barRresolve = \mathbb{E}[ \Rresolve(L) | L>0 ] = \sum_{L=1}^M \frac{L}{S(L)}\hat q(L).
\end{equation}
Further, we are interested in the effective number of bits that is transmitted across the channel per channel use (\ie net rate), denoted by $\barRnet(L)$.
Taking into account that $L$ users each transmit $D$ bits in a total of $S(L)$ slots that each consist of $N$ channel uses we have
\begin{equation}
 \Rnet(L) = \frac{LD}{S(L)N}  = \Rresolve(L)\frac{D}{N}.
\end{equation}
Finally, we are interested in the net rate averaged over $L$, \ie  $\barRnet = \mathbb{E}[ \Rnet(L) | L>0 ]$.

%%%%%%%%%%%%%%%%%%%%%%%%%%%%%%%%%%%%%%%%%%%%%%%%%%%%%%
%
% Preliminaries
%
%%%%%%%%%%%%%%%%%%%%%%%%%%%%%%%%%%%%%%%%%%%%%%%%%%%%%%
\section{Preliminaries} \label{sec:prelim}
%%%%%%%%%%%%%%%%%%%%%%%%%%%%%%%%%%%%%%%%%%%%%%%%%%%%%%
%
% Preliminaries
%
%%%%%%%%%%%%%%%%%%%%%%%%%%%%%%%%%%%%%%%%%%%%%%%%%%%%%%

%%%%%%%%%%%%%%%%%%%%%%%%%%%%%%%%%%%%%%%%%%%%%%%%%%%%%%
\subsection{Signature Codes} \label{ssec:signaturecodes}
% !TEX root =  signatures_main.tex

We are interested in signature coding for the multiple access adder channel with $q$-ary inputs and additions over integers, when up to $K$ random users, out of total $M$ users, are active. There has been a lot of work investigating the case when the signature symbols are binary, \ie $q=2$. A summary of the known asymptotic results has been presented in \cite{MACbook}. However, the case of general $q$ has been significantly less studied.
 
In this paper, we adopt Lindstr\"om's signature coding construction, as presented in \cite[pp. 42 - 43]{MACbook}. The construction is designed for the case that the number of users $M$ is a prime number; if $M$ is not prime, one could design signatures for the smallest prime number larger than $M$ and use just $M$ signatures. For easier exposition, we have assumed in Section~\ref{sec:model} that $M$ is prime.
The construction is performed as follows: Choose integers $s_i$, $i=1,...,M$ such that
\begin{align}
\label{eq:fields}
a^{s_i} = a + b_i, \; i=1,...,M,
\end{align}
where $a$ is a primitive element of $\mathbb{F}_{M^{K}}$ and $b_i$,  $i=1,...,M$, are elements of $\mathbb{F}_{M}$.\footnote{More precisely, in \eqref{eq:fields} we assume summation in $\mathbb{F}_{M^{K}}$, where, with a slight abuse of notation, $b_i$, $i=1,...,M$ also denote corresponding elements from $\mathbb{F}_{M^{K}}$.}
It can be shown that: (i) integers $s_i$, $i=1,...,M$, exist, (ii) $0 < s_i < M^{K} - 1$ and, most importantly, (iii) the sums of subsets of $s_i$ of at most cardinality $K$ have unique values, \ie
\begin{align}
\label{eq:ud}
\sum_{i\in U_1} s_i \neq \sum_{i\in U_2} s_i,  
\end{align}
for any $U_1,U_2\subset\{1,\dots,M\}$, $|U_1|\leq K$, $|U_2|\leq K$ and $U_1\neq U_2$.

The signature of user $i$ is denoted by $W_i^s$. It is a sequence of symbols taking values from $\{0,\dots,q-1\}$ in which the first symbol has value $1$ and the remaining symbols are the $q$-ary representation of the integer $s_i$. Recall from Section~\ref{sec:intro} that the receiver will be dealing with the symbol-wise addition (over the integers) of signatures. Since the first symbol is $1$ for all users the receiver can immediately detect how many users are active and, therefore, determine whether the sum of signatures $s_i$ is uniquely decodable. It is shown in~\cite{MACbook} that in that case also the symbolwise sum of the $W_i^{s}$ is uniquely decodable.  The number of $q$-ary symbols in the above signature code is
\begin{equation}
 \lceil \log_q ( M^K - 1 ) \rceil + 1 \leq K\log_q M + 2.
\end{equation}
Since all rates in this paper are expressed in terms of bits per channel use we express the length of $W_i^s$ in terms of the equivalent number of bits. Let $N_w$ denote the length of $W_i^s$ in terms of bits. We have
\begin{align}
N_w \leq \log_2  \left( K\log_q M + 2 \right) \leq (K+2)\log_2 M,
\end{align}
which holds if $q\leq M$.
 From above, we have the following result.
\begin{theorem}[\cite{MACbook}. pp. 43] \label{th:sigcode}
If $q\leq M$ there exist $q$-ary $K$-out-of-$M$ signature codes that satisfy  
\begin{align}
N_w \leq (K+2)\log_2 M.
\end{align}
\end{theorem}

%%%%%%%%%%%%%%%%%%%%%%%%%%%%%%%%%%%%%%%%%%%%%%%%%%%%%%
\subsection{Reliable Physical-Layer Network Coding} \label{ssec:plnc}
% !TEX root =  signatures_main.tex

Another key ingredient of the random access strategy that is proposed in this paper is to employ physical-layer network coding (PLNC), \ie to organize the physical layer in such a way that the receiver can reliably decode sums of messages that are simultaneously transmitted by users. This requires a suitable choice of the forward error correcting codes as well as the decoding mechanism that is used by the receiver. In this section we provide a short introduction to physical-layer network coding and a result from~\cite{nazer11compforw} that will be needed later. There are various angles at which physical-layer network coding be approached, for instance denoise-and-forward~\cite{popovski2007physical} or compute-and-forward~\cite{nazer11compforw}. A survey of these and other approaches is given in~\cite{nazer2011procieee} and~\cite{Liew2013survey}. In this paper we adopt the compute-and-forward framework as developed by Nazer and Gastpar in~\cite{nazer11compforw}.

\begin{figure}
\centering
\begin{tikzpicture}
\tikzstyle{boxa}=[draw, rounded corners,minimum width=12mm,minimum height=8mm];
\tikzstyle{boxb}=[draw, circle];
\tikzstyle{->}=[-latex];
\tikzstyle{edgea}=[->];
\tikzstyle{edgeb}=[->,olive];

\node[boxa,anchor=east] (encb) at (-1.5,1.5) {\ft{PLNC Encoder $1$}}; 
\node[boxa,anchor=east] (enca) at (-1.5,-1.5) {\ft{PLNC Encoder $2$}}; 

\node[boxa,minimum height=20mm, text width=15mm, align=center] (channel) at (0,0) {\ft{AWGN Multi-access Channel}};
\draw[edgea] (enca) -- (channel); %node[above,near start] {\fm{X_1}}
\draw[edgea] (encb) to  (channel); % node[above,near start] {\fm{X_2}}

\node[boxa,anchor=west] (dec) at (1.5,0) {\ft{PLNC Decoder}};
\draw[edgea] (channel) to (dec); % node[above,near end] {\fm{Y}} 

\draw[edgea] (dec) -- ($ (dec.east) + (8mm,0) $);
\draw[edgea] ($ (enca.west) - (8mm,0) $) -- (enca);
\draw[edgea] ($ (encb.west) - (8mm,0) $) -- (encb);

\node (addermac) at (0,2.5) {\ft{Noiseless $\FF_q$ adder channel}};
\draw[rounded corners] (-3.8,-2.2) rectangle (3.5,2.8);

\end{tikzpicture}
\caption{Physical-layer network coding (PLNC) results in a noiseless $\FF_q$ adder channel. ($K=2$ users)} \label{fig:plnc}
\end{figure}
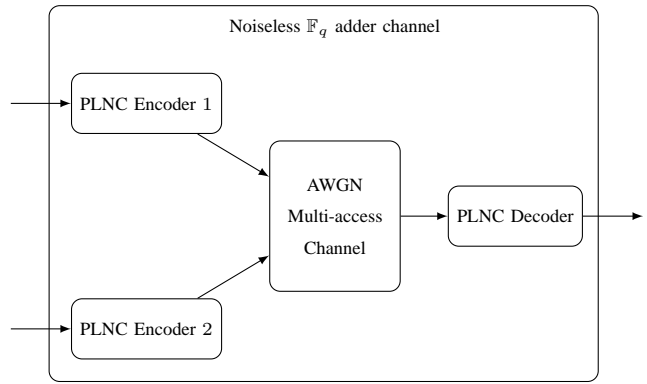

In order to formulate the result from~\cite{nazer11compforw} that we need in the remainder we consider an arbitrary number of $L$ transmitters. User $\ell$ has data $W_\ell$ to transmit, where 
\begin{align}
  W_\ell & = ( W_\ell(1), W_\ell(2), \ldots, W_\ell(\kappa) ),
\end{align}
with $W_\ell(j)\in\mathbb{F}_q$, $q$ prime. Each transmitter uses the same linear code $F$ to encode the data into real-valued channel input of length $N$ (\ie the length of a slot) that satisfies an average power constraint $P$. Let $X_\ell = F(W_\ell)$ denote the channel input of user $\ell$. The decoder, upon observing $Y=\sum_{\ell=1}^L X_\ell + Z$ attempts to decode $(\bigoplus_\ell W_\ell(1), \dots, \bigoplus_\ell W_\ell(\kappa)) $, where $\bigoplus$ denotes adition in $\mathbb{F}_q$. In this sense, the receiver recovers a function (namely, the sum) of the original messages, which is why this approach is referred to as computation coding. In a sense, as illustrated in Figure~\ref{fig:plnc}, we turn the AWGN channel in a noiseless $\FF_q$ adder channel. 

We denote by $\Rplnc$ the rate of $F$, \ie $\Rplnc = \kappa N^{-1} \log_2 q$ bits per channel use. We will refer to $\Rplnc$ as the computation rate and say that it is achievable if the probability of decoding erroneously can be made arbitrarily small by increasing $n$. The next result follows directly from the main result in~\cite{nazer11compforw}. We use the notation
\begin{equation*}
\log^+_2(x) =
\begin{cases}
\log_2(x),\quad &\text{if }x\geq 1, \\
0,\quad &\text{otherwise.} 
\end{cases}
\end{equation*}
\begin{theorem}[\cite{nazer11compforw}, Theorem~1]\label{th:plnc}
For the standard AWGN multiple-access channel the following computation rate is achievable:
\begin{equation}\label{eq:AcheivableComputationRate}
  \Rplnc = \frac{1}{2} \log_2^+ \left( P \right).
\end{equation}
\end{theorem}
The above result does not exactly match the achievable rate as given in\cite[Theorem~1]{nazer11compforw}, which is $\frac{1}{2} \log_2^+ \left( \frac{1}{L} + P \right)$. Since we will be dealing with an unknown number of active users, we use a lower bound on the computation rate, obtained for $L \rightarrow \infty$, which is valid for any number of active users.

The signature codes that we introduced in Section~\ref{ssec:signaturecodes} operate over the adder channel with $q$-ary inputs and additions over the integers. It is important to note that the computation code as described above does not provide an adder channel, but instead provides additions in the finite field $\FF_q$. Therefore, we need an additional result that enables us to lift the computation code result to an integer adder channel. Such a result is provided by Nazer in~\cite{nazer2012successive}. The result states that once the receiver has successfully decoded $(\bigoplus_\ell W_\ell(1), \dots, \bigoplus_\ell W_\ell(\kappa) ) $, \ie the sum over $\FF_q$, it is also possible to recover the integer sum. To state the result more precisely, we create a mapping between the elements of $\FF_q$ and the integers $\{0,1,\dots,q-1\}$. Since we consider $q$ prime such a mapping is trivial. Indeed, 
additions in $\FF_q$ are $\mathrm{mod}\ q$ operations and the elements of $\FF_q$ are naturally identified with the integers $\{0,1,\dots,q-1\}$. With slight abuse of notation we denote by $(\sum_\ell W_\ell(1), \dots, \sum_\ell W_\ell(\kappa))$ the sums of the integers that are identified with the $\FF_q$ elements $W_\ell(k)$. It was shown in~\cite{nazer2012successive} that at $\Rplnc$ as given in Theorem~\ref{th:plnc} the receiver can retrieve the integer sum $\sum_\ell W_\ell$.

%%%%%%%%%%%%%%%%%%%%%%%%%%%%%%%%%%%%%%%%%%%%%%%%%%%%%%
\subsection{Tree Splitting} \label{ssec:treesplitting}

We briefly outline the basic binary tree-splitting algorithm under a collision model~\cite{capetanakis}.
Let $\LL$ denote the set of active users and $L=|\LL|$, $1\leq L\leq M$, denote the number of active users.
In the first slot all $L$ users transmit a packet. If $L = 1$ the receiver successfully decodes the packet of the user and the contention period ends.
If $L\geq 2$ a collision occurs and the receiver does not obtain any useful information. The users probabilistically split into two groups $\LL_{1}$ and $\LL_{2}$. The splitting is uniform at random and independent over users, \ie each user flips a fair coin to decide on the group to join.
Both groups then contend for the medium in the same fashion: first the users from $\LL_1$, then the users from $\LL_2$.
The splitting is done recursively, eventually leading to an instance in which only a single user is active and her transmission is successfully received.
The algorithm continues until the transmissions of all active users from $\LL$ are successfully received.
By means of feedback after each slot the receiver informs the users whether there was a collision, a single or no transmission present, directing the future actions of the contending users.

The above described tree splitting and its variations were thoroughly analyzed in the literature, assessing the performance parameters such as throughput, delay, and stability. The work closest to ours is presented in~\cite{Massey}, the most important difference being that we consider a generalization in which the collision occurs if $L > K$, where $K \geq 1$.
In other words, the use of signatures in the proposed protocol allows for direct exploitation of the slots containing up to $K$ user transmissions, and not just singleton slots.  
The related analysis, which also covers the special case $K=1$, is presented in Section V.

%%%%%%%%%%%%%%%%%%%%%%%%%%%%%%%%%%%%%%%%%%%%%%%%%%%%%%
%
%
%
%%%%%%%%%%%%%%%%%%%%%%%%%%%%%%%%%%%%%%%%%%%%%%%%%%%%%%
\section{Proposed Strategy} \label{sec:idea}
% !TEX root =  signatures_main.tex

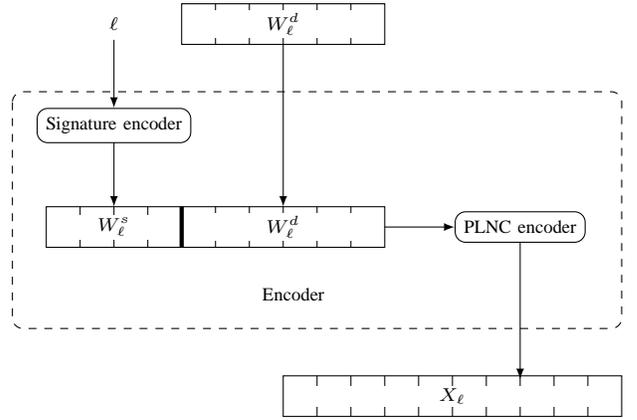
\begin{figure}
\centering
\begin{tikzpicture}[scale=.45]
\tikzstyle{H}=[draw, rectangle,minimum width=6mm,minimum height=6mm];
\tikzstyle{boxb}=[draw, circle,inner sep=1pt];
\tikzstyle{boxc}=[fill=white, circle,minimum width=5mm];
\tikzstyle{->}=[-latex];

\node (ell) at (-2,1) {\fm{\ell}};

\node[draw,rounded corners] (sigcode) at (-2,-2) {\ft{Signature encoder}};

\begin{scope}[xshift=0cm,yshift=1cm]
\draw (0,-.6) rectangle (6,.6);
\foreach \x in {1, 2, ..., 5}
 {
    \draw (\x,-.6) -- (\x,.6);
}
\node[fill=white,inner sep=1pt,minimum width=20mm] at (3,0) {\fm{W_\ell^d}};
\end{scope}

\begin{scope}[xshift=-4cm,yshift=-5cm]
\draw (0,-.6) rectangle (10,.6);
\foreach \x in {1, 2, ..., 9}
 {
    \draw (\x,-.6) -- (\x,.6);
}
\draw[ultra thick] (4,-.6) -- (4,.6);
\node[fill=white,inner sep=1pt,minimum width=10mm] at (2,0) {\fm{W_\ell^s}};
\node[fill=white,inner sep=1pt,minimum width=20mm] at (7,0) {\fm{W_\ell^d}};
\end{scope}

\node[draw,rounded corners] (plnc) at (10,-5) {\ft{PLNC encoder}};

\begin{scope}[xshift=3cm,yshift=-10cm]
\draw (0,-.6) rectangle (10,.6);
\foreach \x in {1, 2, ..., 9}
 {
    \draw (\x,-.6) -- (\x,.6);
}
\node[fill=white,inner sep=1pt,minimum width=40mm] at (5,0) {\fm{X_\ell}};
\end{scope}

\draw[rounded corners,dashed] (-5,-1) rectangle (13,-8);
\node at (3.3,-7) {\ft{Encoder}};

\draw[->] (ell) -- (sigcode);
\draw[->] (-2,-2.5) -- (-2,.-4.4);
\draw[->] (3,.4) -- (3,-4.4);
\draw[->] (6,-5) -- (plnc);
\draw[->] (plnc) -- (10,-9.5);

\end{tikzpicture}
\caption{Illustration of the encoder for user $\ell$ in one slot.\label{fig:encoder}}
\end{figure}

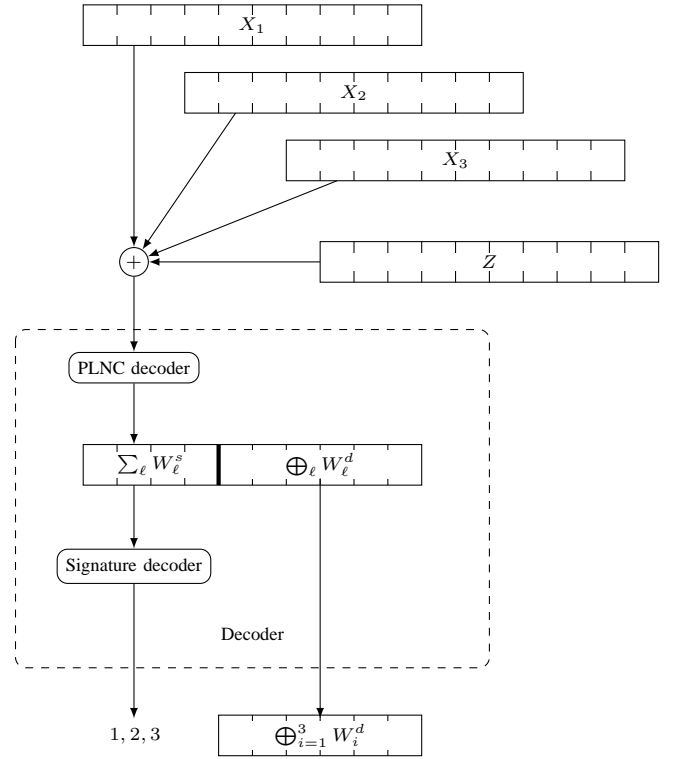
\begin{figure}
\centering
\begin{tikzpicture}[scale=.45]
\tikzstyle{H}=[draw, rectangle,minimum width=6mm,minimum height=6mm];
\tikzstyle{boxb}=[draw, circle,inner sep=1pt];
\tikzstyle{boxc}=[fill=white, circle,minimum width=5mm];
\tikzstyle{->}=[-latex];

\begin{scope}[xshift=0cm,yshift=7cm]
\draw (0,-.6) rectangle (10,.6);
\foreach \x in {1, 2, ..., 9}
 {
    \draw (\x,-.6) -- (\x,.6);
}
\node[fill=white,inner sep=1pt,minimum width=40mm] at (5,0) {\fm{X_{1}}};
\end{scope}

\begin{scope}[xshift=3cm,yshift=5cm]
\draw (0,-.6) rectangle (10,.6);
\foreach \x in {1, 2, ..., 9}
 {
    \draw (\x,-.6) -- (\x,.6);
}
\node[fill=white,inner sep=1pt,minimum width=40mm] at (5,0) {\fm{X_{2}}};
\end{scope}

\begin{scope}[xshift=6cm,yshift=3cm]
\draw (0,-.6) rectangle (10,.6);
\foreach \x in {1, 2, ..., 9}
 {
    \draw (\x,-.6) -- (\x,.6);
}
\node[fill=white,inner sep=1pt,minimum width=40mm] at (5,0) {\fm{X_{3}}};
\end{scope}

\begin{scope}[xshift=7cm,yshift=0cm]
\draw (0,-.6) rectangle (10,.6);
\foreach \x in {1, 2, ..., 9}
 {
    \draw (\x,-.6) -- (\x,.6);
}
\node[fill=white,inner sep=1pt,minimum width=40mm] at (5,0) {\fm{Z}};
\end{scope}

\node[boxb]  (pl) at (1.5,0) {\fm{+}};

\node[draw,rounded corners,anchor=north] (plnc) [below=10mm of pl] {\ft{PLNC decoder}};

\begin{scope}[xshift=0cm,yshift=-6cm]
\draw (0,-.6) rectangle (10,.6);
\foreach \x in {1, 2, ..., 9}
 {
    \draw (\x,-.6) -- (\x,.6);
}
\draw[ultra thick] (4,-.6) -- (4,.6);
\node[fill=white,inner sep=1pt,minimum width=10mm] at (2,0) {\fm{\sum_\ell W^s_\ell}};
\node[fill=white,inner sep=1pt,minimum width=20mm] at (7,0) {\fm{\bigoplus_\ell W^d_\ell}};
\end{scope}

\node[draw,rounded corners] (sigcode) at (1.5,-9) {\ft{Signature decoder}};

\begin{scope}[xshift=4cm,yshift=-14cm]
\draw (0,-.6) rectangle (6,.6);
\foreach \x in {1, 2, ..., 5}
 {
    \draw (\x,-.6) -- (\x,.6);
}
\node[fill=white,inner sep=1pt,minimum width=20mm] at (3,0) {\fm{\bigoplus_{i=1}^3 W^d_{i}}};
\end{scope}

\node[anchor=west] (ell) at (.5,-14) {\fm{1,2,3}};

\draw[rounded corners,dashed] (-2,-2) rectangle (12,-12);
\node at (5,-11) {\ft{Decoder}};

\draw[->] (1.5,6.4) -- (pl);
\draw[->] (4.5,4.4) -- (pl);
\draw[->] (7.5,2.4) -- (pl);
\draw[->] (7,0) -- (pl);
\draw[->] (pl) -- (plnc);
\draw[->] (plnc) -- (1.5,-5.4);
\draw[->] (7,-6.4) -- (7,-13.5);
\draw[->] (1.5,-6.6) -- (sigcode);
\draw[->] (sigcode) -- (1.5,-13.5);

\end{tikzpicture}
\caption{Illustration of the decoder in one slot. ($L=3$ users). \label{fig:decoder}}
\end{figure}

%%%%%%%%%%%%%%%%%%%%%%%%%%%%%%%%%%%%%%%%%%%%%%%%%%%%%%
%
% Idea
%
%%%%%%%%%%%%%%%%%%%%%%%%%%%%%%%%%%%%%%%%%%%%%%%%%%%%%%

We start with an overview of the proposed random access strategy.
The strategy operates in rounds; in each round, the active users transmit the PLNC encoded concatenation of their signatures and payloads.
Use of PLNC enables the receiver to reliably obtain the $q-$ary sums of the user transmissions,
As long as there are at most $K$ active users, the receiver is able to uniquely decode their signatures, detect which users are active and exploit this
information to direct the active users towards solving the linear combination of their payloads.

The receiver is also able to detect when more than $K$ users are active. As explained in Section~\ref{ssec:signaturecodes} this is enabled by the first symbol of the signature that is fixed to one for each user. By observing the integer sum that is decoded in this position the receiver directly learns the number of active users. If this number is larger than $K$, 
the receiver instructs the users to randomly split in two groups and the strategy is then executed in a recursive fashion for each of these groups.
We proceed by presentation of the details.

%%%%%%%%%%%%%%%%%%%%%%%%%%%%%%%%%%%%%%%%%%%%%%%%%%%%%%
\subsection{Encoder}

Referring to Figure~\ref{fig:encoder}, let $W_\ell^s$ and $W_\ell^d$ denote the strings representing the signature and the data payload, respectively, of the active user $\ell$.
According to Theorem~\ref{th:sigcode}, the number of bits in a signature is not more than $(K+2)\log_2 M$.  
The concatenation of signature and payload $W_\ell = W_\ell^s \| W_\ell^d$ is used as the input of a PLNC encoder.
Recall from Section~\ref{ssec:plnc} that the PLNC encoder applies a linear forward error correcting code, the same code $F$ for all users.
The output of the PLNC encoder, denoted by $X_\ell = F(W_\ell) = F(W_\ell^s \| W_\ell^d)$, is a channel input of user $\ell$.

%%%%%%%%%%%%%%%%%%%%%%%%%%%%%%%%%%%%%%%%%%%%%%%%%%%%%%
\subsection{Decoder}

The decoding operation is illustrated in Figure~\ref{fig:decoder}. The receiver observes $Y$, which is a \emph{real} sum 
sum of $X_\ell$, $\ell\in\LL$ and additive noise $Z$,
\begin{equation}
 \sum_{\ell\in\LL} F(W_\ell) + Z.
\end{equation}
It uses a PLNC decoder to decode $Y$ and obtain
\begin{equation}
\bigoplus_{\ell\in\LL} W_\ell, 
\end{equation}
which decomposes into the sums of the signatures $\bigoplus_{\ell\in\LL} W^s_\ell$ and the sums of the codewords $\bigoplus_{\ell\in\LL} W_\ell^d$. Recall from Section~\ref{ssec:plnc} that, once we have obtained the sum $\sum_{\ell\in\LL} W^s_\ell$ over the finite field $\FF_q$, we can also interpret the elements of $W_\ell$ as integers and recover the integer sum $\sum_{\ell\in\LL} W^s_\ell$.

Since the first symbol in the signature of all users is $1$, we directly obtain $L=|\LL|$, the number of active users, from the first symbol in $\sum_{\ell\in\LL} W^s_\ell$. If $L\leq K$, by the property of the signature code we obtain $\LL$ itself.
If $L>K$ no information about $\LL$ can be obtained in this round. 

%%%%%%%%%%%%%%%%%%%%%%%%%%%%%%%%%%%%%%%%%%%%%%%%%%%%%%
\subsection{User Resolution for $L\leq K$}

If $L = |\LL| \leq K$ the receiver has exact knowledge of $\LL$.
Moreover, it has received the sum of the messages $\sum_{\ell\in\LL} W_\ell^d$.
By making use of the feedback mechanism to the users, the receiver ensures that in the next $L-1$ rounds $L-1$ of the users in $\LL$ are individually transmitting their messages. This can be achieved by, for instance, signalling the identity of one the users in the feedback at the end of a round.
In that case the feedback acts as an ACK as well as a scheduling mechanism.

%%%%%%%%%%%%%%%%%%%%%%%%%%%%%%%%%%%%%%%%%%%%%%%%%%%%%%
\subsection{User Resolution for $L>K$}

In case $L > K$ the receiver signals this fact via feedback.
All users in $\LL$ now participate in a splitting protocol with uniform splits into two groups. Each user independently of the other users draws a uniformly distributed random number from $\{1,2\}$. All users with value $1$ enter a new contention resolution phase. The users with value $2$ wait until this phase ends and start another contention resolution phase afterwards. If there are more than $K$ users in one of these groups the splitting procedure is applied recursively.

In the next section we analyze the proposed strategy, parametrized on the values of $K$.
Note that case $K =1$ reduces the scheme to the traditional tree splitting protocol that was discussed in Section~\ref{ssec:treesplitting}.

%%%%%%%%%%%%%%%%%%%%%%%%%%%%%%%%%%%%%%%%%%%%%%%%%%%%%%
%
% Basic strategy
%
%%%%%%%%%%%%%%%%%%%%%%%%%%%%%%%%%%%%%%%%%%%%%%%%%%%%%%
\section{Analysis} \label{sec:analysis}
% !TEX root =  signatures_main.tex

%%%%%%%%%%%%%%%%%%%%%%%%%%%%%%%%%%%%%%%%%%%%%%%%%%%%%%
%
% Analysis of the basic strategy
%
%%%%%%%%%%%%%%%%%%%%%%%%%%%%%%%%%%%%%%%%%%%%%%%%%%%%%%

%%%%%%%%%%%%%%%%%%%%%%%%%%%%%%%%%%%%%%%%%%%%%%%%%%%%%%
\subsection{User Resolution Rate}

%%%
%
% Figure
%
%%%
\begin{figure}
\centering
\begin{tikzpicture}
\begin{axis}[
  xlabel=$L$,ylabel=$S$, %\rotatebox{-90}
  xmin=0, xmax=20,
  ymin=0,ymax=30,
  font=\scriptsize,
  legend style={
        cells={anchor=west},
        legend pos=south east,
       font=\scriptsize,
    }
]

\addplot[
  line width=.3mm,color=blue, solid,
  mark=square*,mark repeat=10,mark phase=0,mark size=.5mm,mark options={solid}
  ]
table[
  header=false,x index=0,y index=1,
  ]
{matlab_figures/SinL.csv};
\addlegendentry{$K=1$};
\addplot[
  line width=.3mm,color=blue, dashed,
 mark=square*,mark repeat=10,mark phase=0,mark size=.5mm,mark options={solid},
 forget plot
  ]
table[
  header=false,x index=0,y index=2,
  ]
{matlab_figures/SinL.csv};
\addplot[
  line width=.3mm,color=blue, dotted,
 mark=square*,mark repeat=10,mark phase=0,mark size=.5mm,mark options={solid},
 forget plot
  ]
table[
  header=false,x index=0,y index=3,
  ]
{matlab_figures/SinL.csv};
\addplot[
  line width=.3mm,color=red, solid,
  mark=square,mark repeat=10,mark phase=0,mark size=.5mm,mark options={solid}
  ]
table[
  header=false,x index=0,y index=4,
  ]
{matlab_figures/SinL.csv};
\addlegendentry{$K=4$};
\addplot[
  line width=.3mm,color=red, dashed,
  mark=square,mark repeat=10,mark phase=0,mark size=.5mm,mark options={solid},
forget plot
  ]
table[
  header=false,x index=0,y index=5,
  ]
{matlab_figures/SinL.csv};
\addplot[
  line width=.3mm,color=red, dotted,
  mark=square,mark repeat=10,mark phase=0,mark size=.5mm,mark options={solid},
forget plot
  ]
table[
  header=false,x index=0,y index=6,
  ]
{matlab_figures/SinL.csv};
\addplot[
  line width=.3mm,color=green, solid,
  mark=*,mark repeat=10,mark phase=0,mark size=.5mm,mark options={solid}
  ]
table[
  header=false,x index=0,y index=7,
  ]
{matlab_figures/SinL.csv};
\addlegendentry{$K=16$};
\addplot[
  line width=.3mm,color=green, dashed,
  mark=*,mark repeat=10,mark phase=0,mark size=.5mm,mark options={solid},
 forget plot
  ]
table[
  header=false,x index=0,y index=8,
  ]
{matlab_figures/SinL.csv};
\addplot[
  line width=.3mm,color=green, dotted,
  mark=*,mark repeat=10,mark phase=0,mark size=.5mm,mark options={solid},
 forget plot
  ]
table[
  header=false,x index=0,y index=9,
  ]
{matlab_figures/SinL.csv};
\end{axis}
\end{tikzpicture}
\caption{$S(L)$ and its bounds for various values of $K$. Upper and lower bounds in dotted and dashed lines, respectively. Exact values of $S(L)$ in solid lines.\label{fig:SinL}}
\end{figure}
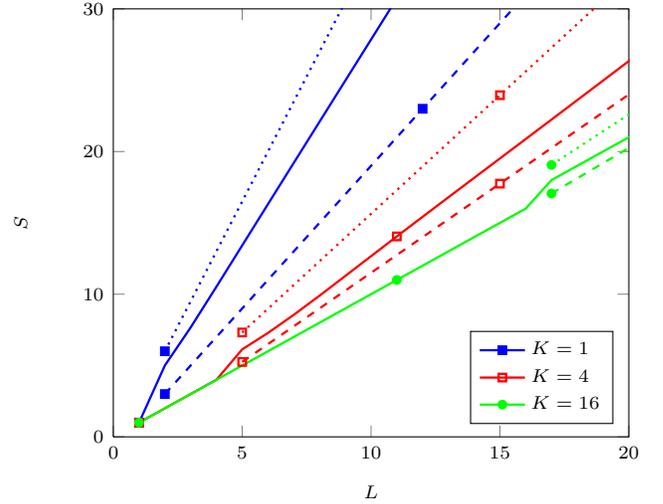

%\begin{table}
%\centering
%\begin{equation*}
%\begin{array}{rll}
%K & \alpha^* & \beta^* \\
%\hline
%1 &  2         & 3.5 \\
%2 &  1.5      & 2.278 \\
%4 &  1.25    & 1.663 \\
%8 &  1.125  & 1.348 \\
%16 & 1.063 & 1.18
%\end{array}
%\end{equation*}
%\caption{Values for $\alpha^*$ and $\beta^*$ that serve in the bounds on $S(L)$.} \label{table:TLbounds}
%\end{table}

We provide an analysis in terms of a recursive expression for $S(L)$, the expected number of slots in a contention period in terms of the number of active users $L$.
%The analysis is similar to the one by Massey~\cite{Massey}. 
We start by stating the main result of this section; the proof is omitted due to space constraints.
%given in the appendix. 
%Let
\begin{align}
\alpha^* &= 1+\frac{1}{K}, \\
\beta^* &= 1 + \frac{1}{(K+1)(2^K-1)} + \frac{2}{K+1} + \frac{1}{K}.
\end{align}
\begin{theorem} \label{th:boundsbasic}
$S(L) = L $ if $1\leq L\leq K$, and, for $L>K$
\begin{equation}
\alpha^* L - 1 \leq S(L) \leq \beta^* L - 1.
\end{equation}
\end{theorem}
In Figure~\ref{fig:SinL} we have illustrated $S(L)$ as well as the above bounds for various values of $K$.
%In Table~\ref{table:TLbounds} we provide a numerical evaluation of the bounds.

From Theorem~\ref{th:boundsbasic} we derive results on $\barRresolve$, the expected number of users that is resolved per slot.
\begin{theorem} \label{th:Rresolve}
The expected number of users that is resolved per round is lower bounded as
\begin{equation}
\barRresolve \geq \frac{\beta^* I_{1-p}(M\!-\!K,K\!+\!1) + I_{p}(K\!+\!1,M\!-\!K) - q_0\beta^*}{(1-q_0)\beta^*}.
\end{equation}
\end{theorem}
\begin{IEEEproof}
We have
\begin{align}
\barRresolve
&= \sum_{L=1}^M \frac{L}{S(L)}\hat q(L) \\
&\geq \sum_{L=1}^K \hat q(L) + \sum_{L=K+1}^M \frac{L}{\beta^*L-1}\hat q(L) \\
&\geq (1-q_0)^{-1}\left( \sum_{L=0}^K q(L) + \frac{1}{\beta^*}\sum_{L=K+1}^M q(L) - q_0 \right) \\
&= \frac{\beta^* I_{1-p}(M\!-\!K,K\!+\!1) + I_{p}(K\!+\!1,M\!-\!K) - q_0\beta^*}{(1-q_0)\beta^*}.
\end{align}
\end{IEEEproof}

\begin{figure}
\centering
\begin{tikzpicture}
\begin{axis}[
  xlabel=$K$,ylabel=$\barRresolve$, %\rotatebox{-90} ymin=0,ymax=1.5,
  font=\scriptsize,
  legend style={
        cells={anchor=west},
        legend pos=south east,
       font=\scriptsize,
    }
]

\addplot[
  line width=.3mm,color=blue, solid,
  mark=square*,mark size=.5mm,mark options={solid}
  ]
table[
  header=false,x index=0,y index=1,
  ]
{matlab_figures/avgperf_inK.csv};
\addlegendentry{$p=3/M$};
\addplot[
  line width=.3mm,color=red, solid,
  mark=square,mark size=.5mm,mark options={solid}
  ]
table[
  header=false,x index=0,y index=2,
  ]
{matlab_figures/avgperf_inK.csv};
\addlegendentry{$p=6/M$};
\addplot[
  line width=.3mm,color=green, solid,
  mark=*,mark size=.5mm,mark options={solid}
  ]
table[
  header=false,x index=0,y index=3,
  ]
{matlab_figures/avgperf_inK.csv};
\addlegendentry{$p=12/M$};
\end{axis}
\end{tikzpicture}
\caption{Lower bounds on $\barRresolve$, the expected number of users that is resolved per slot.  ($M=1031$)\label{fig:avgperf_inK}}
\end{figure}
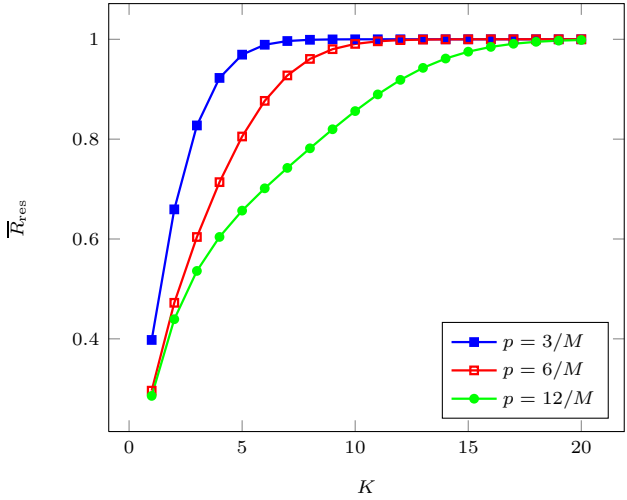

The result is illustrated in Figure~\ref{fig:avgperf_inK} as a function of $K$ for various values of $p$.

%%%
%
% Figure
%
%%%

%%%%%%%%%%%%%%%%%%%%%%%%%%%%%%%%%%%%%%%%%%%%%%%%%%%%%%
\subsection{Rate in Bits per Channel Use}
In the previous subsection we analyzed $\Rresolve(L)$, the expected number of users that is resolved in a slot.
In this section we consider $\Rnet(L)$, which is the overall throughput in bits per channel use that is effectively transmitted. The overall throughput takes into account the overhead that is generated by the signatures as well as the physical-layer network coding.

It is readily verified that from Theorems~\ref{th:sigcode},~\ref{th:plnc} and~\ref{th:Rresolve} it follows that
\begin{equation}
\Rnet(L) \geq \Rresolve(L)\Rplnc\frac{D}{(K+2)\log_2 M + D}.
\end{equation}
 
This leads to the following corollary to Theorem~\ref{th:Rresolve}.
\begin{corollary} \label{cor:Rnet}
The expected number of bits per channel use $\barRnet$ is at least
\begin{multline}
\barRnet \geq \sum_{L=1}^M \binom{M}{L}p^L(1-p)^{M-L} \frac{L}{\beta^*L-1} \\ \frac{\log^+_2(P)}{2}\frac{D}{(K+2)\log_2 M + D}.
\end{multline}
\end{corollary}

Finally, we consider an information-theoretic upper bound on $\Rnet$, \ie an upper bound that must be satisfied by any protocol. The bound is obtained by assuming that the receiver knows which users are active and that these users can employ a multiuser code which is optimaly decoded by the receiver. Under these assumptions the problem reduces to a standard Gaussian multi-access channel. The sum rate that can be used by $L$ active users is
%\begin{equation}
$\Rnet(L) \leq \frac{1}{2}\log_2\left(1+LP\right)$.
%\end{equation}
This immediately leads to
\begin{equation} \label{eq:upper}
\barRnet \leq \sum_{L=1}^M \binom{M}{L}p^L(1-p)^{M-L} \frac{1}{2}\log_2\left(1+LP\right).
\end{equation}

\begin{figure}
\centering
\begin{tikzpicture}
\begin{semilogxaxis}[
  xlabel=$D$ (bits),ylabel=$\barRnet$, %\rotatebox{-90}
  font=\scriptsize,
  legend style={
        cells={anchor=west},
        legend pos=south east,
       font=\scriptsize,
    }
]

\addplot[
  line width=.3mm,color=blue, solid,
  mark=square*,mark repeat=10,mark phase=0,mark size=.5mm,mark options={solid}
  ]
table[
  header=false,x index=0,y index=1,
  ]
{matlab_figures/Rtotal_inD.csv};
\addlegendentry{$K=3$};
\addplot[
  line width=.3mm,color=red, solid,
  mark=square,mark repeat=10,mark phase=0,mark size=.5mm,mark options={solid}
  ]
table[
  header=false,x index=0,y index=2,
  ]
{matlab_figures/Rtotal_inD.csv};
\addlegendentry{$K=8$};
\addplot[
  line width=.3mm,color=green, solid,
  mark=*,mark repeat=10,mark phase=0,mark size=.5mm,mark options={solid}
  ]
table[
  header=false,x index=0,y index=3,
  ]
{matlab_figures/Rtotal_inD.csv};
\addlegendentry{$K=16$};
\addplot[
  line width=.3mm,color=black, dashed,
  forget plot
  ]
table[
  header=false,x index=0,y index=4,
  ]
{matlab_figures/Rtotal_inD.csv};
\addplot[
  line width=.3mm,color=black, dotted,
  forget plot
  ]
table[
  header=false,x index=0,y index=5,
  ]
{matlab_figures/Rtotal_inD.csv};
\end{semilogxaxis}
\end{tikzpicture}
\caption{Lower bounds on $\barRnet$. In dashed line the upper bound on $\barRnet$. In dotted line the value $1/2\log_2(P)$. ($M=1031$, $p=3/M$, $P=10^2$) \label{fig:Rtotal_inD}}
\end{figure}
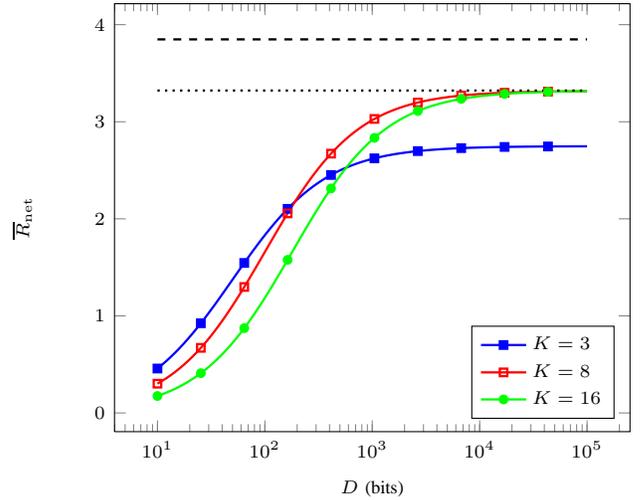

In Figure~\ref{fig:Rtotal_inD} we have illustrated our lower bound on $\barRnet$ as a function of $D$, the size the data packet. In addition, Figure~\ref{fig:Rtotal_inD} illustrates the upper bound~\eqref{eq:upper}.
 
%%%
%
% Figure
%
%%%

%%%%%%%%%%%%%%%%%%%%%%%%%%%%%%%%%%%%%%%%%%%%%%%%%%%%%%
\subsection{Evaluation}
Figure~\ref{fig:avgperf_inK} shows that as a function of $K$, $\barRresolve$ quickly approaches 1. This performance parameter is a baseline measure of the efficiency of the random access protocols from the system perspective and is usually refered to a throughput. Our results clearly demonstrate the potential of the proposed strategy.
Specifically, the proposed strategy outperforms all the state-of-the-art random access schemes in terms of $\barRresolve$, cf. Section~\ref{sec:intro}.
Figure~\ref{fig:avgperf_inK}  also shows that $K$ should increase as the expected number of the active users $pM$ increases, if high throughputs are to be achieved.
Conversely, if $K$ is fixed the expected throughput drops with $p$. This implies that one should design $K$ to match the expected number of active users.
This requirement is similar to the ALOHA-based protocols, where the optimal values of the protocol parameters, like frame lengths or user activation probability, depend on the number of active users \cite{R1975,PSLP2014,CGH2007}.

Fig.~\ref{fig:Rtotal_inD} demonstrates what is the price to pay in information bits per channel use  due to: (i) the overhead related to the use of signatures and (ii) information waste caused by collisions, compared to the idealized solution of aforehand knowing the set of active users and using the optimal multi-user code, and iii) the use of physical-layer network coding.
Obviously, this loss is pronounced for low payload lengths $D$ and diminishes as $D$ increases. Also, it is reflected in Figure~\ref{fig:Rtotal_inD} that for large $D$ the loss that is incurred is only due to the physical-layer network coding. It is 
%currently
 an open problem if this loss is an inherent property of physical-layer network coding or an artifact of the computation coding construction that is developed in~\cite{nazer11compforw}.

It is interesting to observe that the depicted results clearly suggest that due to the counter balancing of effects (i) and (ii) there is an optimal choice of $K$ with respect to $D$.
Finally, we note that the state-of-the-art random access protocols in general suffer from the same limitations; e.g., in SIC-based ALOHA solutions one has to invest overhead in pointers to packet replicas.

%%%%%%%%%%%%%%%%%%%%%%%%%%%%%%%%%%%%%%%%%%%%%%%%%%%%%%
%
%
%
%%%%%%%%%%%%%%%%%%%%%%%%%%%%%%%%%%%%%%%%%%%%%%%%%%%%%%
\section{Discussion} \label{sec:discussion}

As a further work, we consider extensions dealing with: errors that occur due physical-layer network coding at finite block lengths, more general user activity models (including their absence, as well), sensitivity of the performance parameters to the choice of $K$ and variations of the scheme when the feedback channel is limited.

%%%%%%%%%%%%%%%%%%%%%%%%%%%%%%%%%%%%%%%%%%%%%%%%%%%%%%
%
%
%
%%%%%%%%%%%%%%%%%%%%%%%%%%%%%%%%%%%%%%%%%%%%%%%%%%%%%%
\section*{Acknowledgement}
This work was supported in part by the Netherlands Organization for Scientific Research (NWO), grant $612.001.107$. The work was supported in part by the Danish Council for Independent Research (DFF), grant no. 11-105159 ``Dependable Wireless Bits for Machine-to-Machine (M2M) Communications'' and grant no. DFF-4005-00281 ``Evolving wireless cellular systems for smart grid communications''.

%%%%%%%%%%%%%%%%%%%%%%%%%%%%%%%%%%%%%%%%%%%%%%%%%%%%%%
%
%
%
%%%%%%%%%%%%%%%%%%%%%%%%%%%%%%%%%%%%%%%%%%%%%%%%%%%%%%
\bibliographystyle{IEEEtran}
\bibliography{IEEEabrv,signatures}

\end{document}